\documentclass[angeo]{copernicus}
\nolinenumbers

\usepackage{subcaption}
\usepackage{graphicx}

\begin{document}

\title{Permutation Entropy and Complexity Analysis of Large-scale Solar Wind Structures and Streams}

\Author[1]{Emilia}{Kilpua}
\Author[1]{Simon}{Good}
\Author[2,1]{Matti}{Ala-Lahti}
\Author[1]{Adnane}{Osmane}
\Author[1]{Venla}{Koikkalainen}

\affil[1]{Department of Physics, University of Helsinki, P.O. Box 64, FI-00014 Helsinki, Finland}
\affil[2]{Department of Climate and Space Sciences and Engineering, University of Michigan, 2455 Hayward St., Ann Arbor, MI 48109-2143, USA}

\correspondence{Emilia Kilpua (emilia.kilpua@helsinki.fi)}

\runningtitle{Permutation Entropy and Complexity of the Solar Wind}

\runningauthor{Kilpua et al.}

\maketitle

\begin{abstract}
In this work, we perform a statistical study of magnetic field fluctuations in the solar wind at 1~au using permutation entropy and complexity analysis, and the investigation of the temporal variations of the Hurst exponents. Slow and fast wind, magnetic clouds, interplanetary coronal mass ejection (ICME)-driven sheath regions and slow--fast stream interaction regions (SIRs) have been investigated separately. Our key finding is that there are significant differences in permutation entropy and complexity values between the solar wind types at larger timescales and little difference at small timescales. Differences become more distinct with increasing timescale, suggesting that smaller-scale turbulent features are more universal. At larger timescales, the analysis method can be used to identify localized spatial structures. We found that except in magnetic clouds, fluctuation are largely anti-persistent and that the Hurst exponents, in particular in compressive structures (sheaths and SIRs) exhibit a clear locality. Our results shows that, in all cases apart from magnetic clouds at largest scales, solar wind fluctuations are stochastic with the fast wind having the highest entropies and low complexities. Magnetic clouds in turn exhibit the lowest entropy and highest complexity, consistent with them being coherent structures in which the magnetic field components vary in an ordered manner. SIRs, slow wind and ICME sheaths are intermediate to magnetic clouds and fast wind, reflecting the increasingly ordered structure. Our results also indicate that permutation entropy -- complexity analysis is a useful tool for characterizing the solar wind and investigating the nature of its fluctuations. 

\end{abstract}

\introduction
\label{sec:introduction}

The study of multi-scale magnetic field fluctuations is an active research area in space, astrophysical and laboratory plasmas. One of the few natural environments in which it is possible to study them with direct measurements is the collisionless solar wind that incessantly streams from the Sun and fills the heliosphere \citep[e.g.,][]{Bruno2013}. Solar wind fluctuations are generally thought to arise from waves, turbulence and coherent structures, but many open questions regarding their nature and evolution prevail. So-called `mesoscale' solar wind structures, corresponding to structures with spatial extents of approximately 5-10,000~Mm and temporal scales ranging from $\sim$ 10~s to 7~h near Earth's orbit ($\sim 1$ AU), have also recently been brought to the centre of attention due to their importance in solar wind formation and evolution, and their space weather impacts \citep[e.g.,][]{Viall2021}.

Outward-propagating incompressible Alfv\'{e}nic fluctuations from the Sun are a common feature of fast solar wind streams \citep[e.g.,][]{Belcher1971}. Their non-linear interaction with inward-directed Alfv\'{e}n waves generated locally in interplanetary space \citep[e.g.][]{Chen2020} are believed to drive a turbulent cascade of energy from large to small scales, where energy finally dissipates and heats the solar wind \citep[][]{Smith2021}. The slow solar wind is also turbulent but with a more variable structure and a higher occurrence of coherent intermittent structures \citep[e.g.,][]{Bruno2003,Wawrzaszek2021}. Knowledge of the properties and nature of magnetic field fluctuations is also important for understanding large-scale heliospheric structures, such as interplanetary coronal mass ejections (ICMEs) and their sheaths \citep[][]{Kilpua2017a}, and slow--fast stream interaction regions \citep[SIR; e.g.,][]{Richardson2018}, as well as for understanding how energy is transferred through their boundaries. In addition, magnetic fluctuations have an important role in the acceleration and transport of solar energetic particles \citep[SEPs; e.g.,][]{Oughton2021}, and highly fluctuating solar wind is also considered more geoeffective \citep[e.g.,][]{Borovsky2003,Osmane2015,Kilpua2017b,Telloni2021,Han2023,Lei2023}.

Permutation entropy analysis \citep{Bandt2002} and Jensen-Shannon complexity analysis \citep{Rosso2007} are powerful tools for investigating fluctuations. They have been used in widely-ranging contexts and also recently in space plasma physics studies \citep[][]{Weck2015,Weygand2019,Osmane2019,Good2020a, Olivier2019, Kilpua2022, Raath2022}. The determination of permutation entropy and Jensen-Shannon complexity is based on investigating the occurrence of permutation (or ordinal) patterns in time series. An ordinal pattern is formed by a set of subsequent values in a time series separated by time lag $\tau$ and it thus gives information on the relation between the values forming the pattern. Varying $\tau$ allows fluctuations over multiple time scales to be investigated. The number of elements in an ordinal pattern is called the embedded dimension, $d$, and the factorial of $d$ gives the number of possible permutations. The frequency at which different ordinal patterns occur in a time series determines its entropy and predictability. For example, if only a few ordinal patterns are present (i.e., all other permutations have zero probability), permutation entropy is close to zero, signifying high predictability or knowledge of the underlying process. A situation in which all ordinal patterns occur with the same probability yields the maximum entropy state, signifying low predictability.

However, permutation entropy cannot yield information about the randomness of the patterns or the structural complexity of time series. Complexity is related to how far the distribution formed by all permutations in the time series is from the maximum entropy (uniform) distribution \citep[e.g.,][]{Zanin2021}. Both highly ordered cases (e.g., periodic fluctuations like sine waves) and random cases (e.g., rolling of a dice, white and pink noise) have low complexities. Note that the former case has lower entropy and the latter close to the maximum entropy. In between the zero and maximum entropy cases, complexity can have a range of values. Maximal complexities are associated with chaotic fluctuations that are structured but have lower predictability.

All those studies made in the solar wind using complexity and entropy analysis so far have indicated that magnetic field fluctuations are stochastic \citep[][]{Weck2015,Weygand2019, Good2020a,Kilpua2022, Raath2022}. \cite{Weck2015} analyzed both laboratory plasmas and solar wind. Their study suggests that solar wind fluctuations represent fully developed turbulence, while fluctuations in the investigated laboratory settings were weakly turbulent or not even truly turbulent. \cite{Weygand2019} investigated the complexity of solar wind magnetic fluctuations in ICMEs and SIRs using both Wind and Ulysses data at distances from 1 to 5.4~au, while \citep{Raath2022} analysed Ulysses observations up to 0.34 au with the focus on periods showing low entropy. It was found that the entropy increased and complexity decreased with the increasing heliospheric distance, suggesting that solar wind fluctuations become more stochastic in nature. \cite{Good2020a} and \citep[][]{Kilpua2022} were both case studies of a slow ICME-driven sheath. 

Another important characterization of the nature of time series is to explore their memory and correlations via the Hurst exponent analysis \citep[e.g.,][]{Ruzmaikin1994}. This approach has also been used in some space physics studies, in particular to characterize geomagnetic field fluctuations and geomagnetic activity \citep[e.g.,][]{Balasis2006,Balasis2023,Michelis2016}. These studies have for example demonstrated that the scaling properties of geomagnetic fluctuation change with with magnetospheric activity levels.

In this paper we investigate the occurrence of ordinal patterns, entropy and complexity in different types of solar wind. The categories include slow and fast solar wind, ICME sheaths, SIRs and magnetic clouds. In addition, we investigate the memory of the time series extracted from these structures by analysing their magnetic field fluctuation scaling properties with the Hurst exponent. 

\section{Methods and approaches}

\subsection{Data and event selections}

We here use 3~s magnetic field data from the Magnetic Fields Investigation \citep[MFI;][]{lepping1995} fluxgate magnetometer on board the \textit{Wind} spacecraft \citep{ogilvie1995}. The data are analyzed in Geocentric Solar Ecliptic (GSE) coordinates. The events are gathered from 1997 to 2022, i.e. spanning over two solar cycles.

Data intervals of 12~h duration were taken from the following types of solar wind, with the number of events for each solar wind type given in parenthesis: 1) slow wind (55), 2) fast wind (49), 3) SIR (70), 5) ICME-driven sheath regions (27), and 7) magnetic clouds (74). The lower number of sheath events is explained by the requirement to have the interval duration at least 12~h. We note that sheaths and SIRs may in particular have some significant variations in their properties over the interval \citep[e.g.,][]{Kilpua2017a}, but we did not separate them into sub-intervals to have as long as possible durations for investigation. In SIRs, the stream interface (SI) separates the cooler, denser and slower solar wind from the tenuous, hot and fast wind \citep[e.g.,][]{Gosling1978, Richardson2018}. In sheaths, the field and plasma close to the shock has been recently processed by the shock, while close to the ICME leading edge plasma and field have evolved considerably since encountering the shock and could have been further modified by processes at the sheath-ICME boundary, e.g. field line draping. Both sheaths and SIRs are compressive structures, but a typical SIR at 1~au is not preceded by a shock \citep{Jian2006}. We here focus on the subset of ICMEs called magnetic clouds, which represent large-scale flux ropes ejected from the Sun \citep[e.g.,][]{Burlaga1981,Klein1982}. 

The magnetic clouds were collected from the \textit{Wind} ICME catalogue  \citep{Nieveschinchilla2018} and the Richardson and Cane ICME list \citep{richardson2010}, the SIRs times from the ACE/WIND Stream Interaction Regions catalog, and sheaths from the list published by \cite{Kilpua2021Fr}. We here considered SIRs where the solar wind speed reached at least 650~km~s$^{-1}$ after the SIR. Fast wind intervals were defined as the periods when the average wind speed during a 12~h interval after the SIR was $>$~600~km~s$^{-1}$. The slow solar wind intervals were defined as the periods preceding the SIR during which the 12~h averaged wind speed was $<$~450~km~s$^{-1}$. 

\subsection{Permutation entropy and Jensen-Shannon complexity}

As discussed in Section~\ref{sec:introduction}, there are $d!$ possible permutations for the embedded dimension $d$. For example, if $d=5$, there are 120 possible permutations: 12345 (indicating that samples in the series have a monotonously ascending order from beginning to end), 13245, 12425, etc. If we denote the probability of permutation $j$ to be $p_j$, and the set of probabilities $P$, the permutation entropy according to \cite{Bandt2002} is defined as
\[ S(P) = -  \sum_{j=1}^{d!} p_j \log{p_j} \]
From above we see that $S(P) = 0$ when only one permutation occurs and it maximizes when all permutations occur with equal probability.

The normalized Shannon permutation entropy
\[  \mathcal{H_S}(P) = S(P)/\log{d!} \]
is defined such that it takes values between 0 and 1, where 0 is the lowest entropy and 1 the maximum entropy.

The Jensen-Shannon complexity as defined by \cite{Rosso2007} is
\[ C_J^S = - 2 \frac{S(\frac{P+P_e}{2})-\frac{1}{2} S(P) - \frac{1}{2} S(P_e)}{\frac{d!+1}{d!} \log{(d!+1)} - 2 \log{(2d!)} + \log{d!}} \mathcal{H_S}(P)  \]
The quantity $S(\frac{P+P_e}{2})-\frac{1}{2} S(P) - \frac{1}{2} S(P_e)$ is the Jensen-Shannon divergence, which is a measure of similarity between two probability distributions. In this case, $P$ is the distribution formed by the patterns in the investigated time series and $P_e$ is the distribution that maximizes the permutation entropy, i.e., the one where all permutations occur with equal probability. As stated in Section~\ref{sec:introduction}, both the perfectly random ($P = P_e$) and perfectly ordered ($\mathcal{H_S}(P) \approx 0$) case yields zero complexity. In general, complexity is lower the closer the distribution $P$ is to the maximum entropy state and the lower the normalized permutation entropy $H(P)$. The highest complexity values require the repeated occurrence of certain patterns that reflect underlying chaotic processes,  i.e. they occur when distribution $P$ is far from $P_e$, but has higher normalized entropy. 

The statistical robustness has been tested for the complexity-entropy analysis. The analyzed time series must be long enough so that enough permutation sequences can be extracted to allow all possible permutations to be sufficiently sampled. The commonly used robustness criteria are  $N/d! > 5$ and $\sqrt{d!/(N-(d-1) r)} < 0.2$  \citep[e.g.,][]{Osmane2019,Weygand2019} where $N$ is the number of samples in the investigated time series and $r$ gives the subsampling rate, i.e. the time lag $\tau = r \Delta t$, where $\Delta t$ is the data resolution. Here the length of each time series is 12~h at 3~s resolution, thus giving 14,000 samples in each interval. We here use $d=5$, similar to many previous studies of the solar wind \citep{Weck2015, Weygand2019,Good2020a}.
The embedded dimension of 5 gives $N/d! = 120$ and we limit our analysis to $r = 600$ ($\tau =$ 1800~s). This gives $\sqrt{d!/(N-(d-1) r)} = 0.10$, indicating that both robustness criteria are well met. 

\subsection{Hurst exponent}
\label{sec:hurst_explanation}

We also calculate the Hurst exponents for the investigated time series. As mentioned in the Introduction, the Hurst exponent, $H$, is used to characterize the memory and correlations in stochastic time series. We calculate here the Hurst exponent from the first-order structure function that for a time  series $x(t)$ is defined as \citep[e.g.,][]{Matteo2007,Michelis2016,Gilmore2002,Giannattasio2022} 
$$
S_1(\tau) = \langle | x(t + \tau) - x(t) | \rangle \sim \tau^{H}  
$$
When the scaling is linear, the Hurst exponent can be estimated as linear regression fits to $log_{10}(\langle | x(t + \tau) - x(t) | \rangle $) as a function of $log_{10}(\tau$).  In practice, this means calculating auto-correlations of $x$ with time lags $\tau$. 
 
The value $H \sim 0.5$ signifies an uncorrelated random walk with short-memory (general Brownian motion or Brown noise) where the mean-squared distance from the starting point of the walk increases with time. Such a time series is uncorrelated in the sense that the steps in the random walk are independent of each other or, in other words, that the auto-correlation of the time series is zero. Hurst exponent values $< 0.5$ refer to mean-reverting time series where increases are followed by decreases and vice verse, while time series $H > 0.5$ are said to be persistent where increase is followed by an increase or decrease by another decrease.

For time series that are monofractal, the Hurst exponent relates to the spectral slope $\alpha$ of the fluctuation power spectrum $f^{-\alpha}$ as $\alpha = 2H + 1$ \citep{Mandelbrot1977,Matteo2007}. This gives $H = 0.33$ for the Kolmogorov scaling with spectral slope $\alpha = -5/3$ \citep{kolmogorov1941} and $H = 0.25$ for the Iroshnikov-Kraichnan scaling with spectral slope $\alpha -3/2$ \citep[][]{iroshnikov1964,kraichnan1965}. There is however indications that solar wind time series are generally multi-fractal in nature \citep[e.g.,][]{Marsch1997,Bruno2019MF,Gomes2023,Kilpua2020}, and therefore, this relation needs to be considered with caution.

As the scaling properties of the time series may not be constant we use a local Hurst exponent estimated by sliding a time window through the analysed time series, as was done for example by \cite{Michelis2016}. The authors note that the window length needs to be at least ten times larger than the maximum scale investigated. We here use a window width of 18,000~s (5~h) sliding in 30~min steps through each 12~h time series that allows to have the largest investigated time lag as 1800~s (30~min). The Hurst exponents are determined over 360~s (6~min) wide time lag intervals ($\Delta \tau$) in $\tau =30$~s steps from 60~s to 1740~s. These 360~s intervals are stepped forward in 90~s steps. The events where the standard error in the fitting was $> 0.05$ were removed from the analysis. We however note that there are now drastic differences in the results when all exponents are included (data not shown), suggesting that removal of the points does not have strong influence. 

\section{Result}
\label{sec:results}

\subsection{Examples}
\label{sec:examples}

\begin{figure}[ht]
\centering
\includegraphics[width=\linewidth]{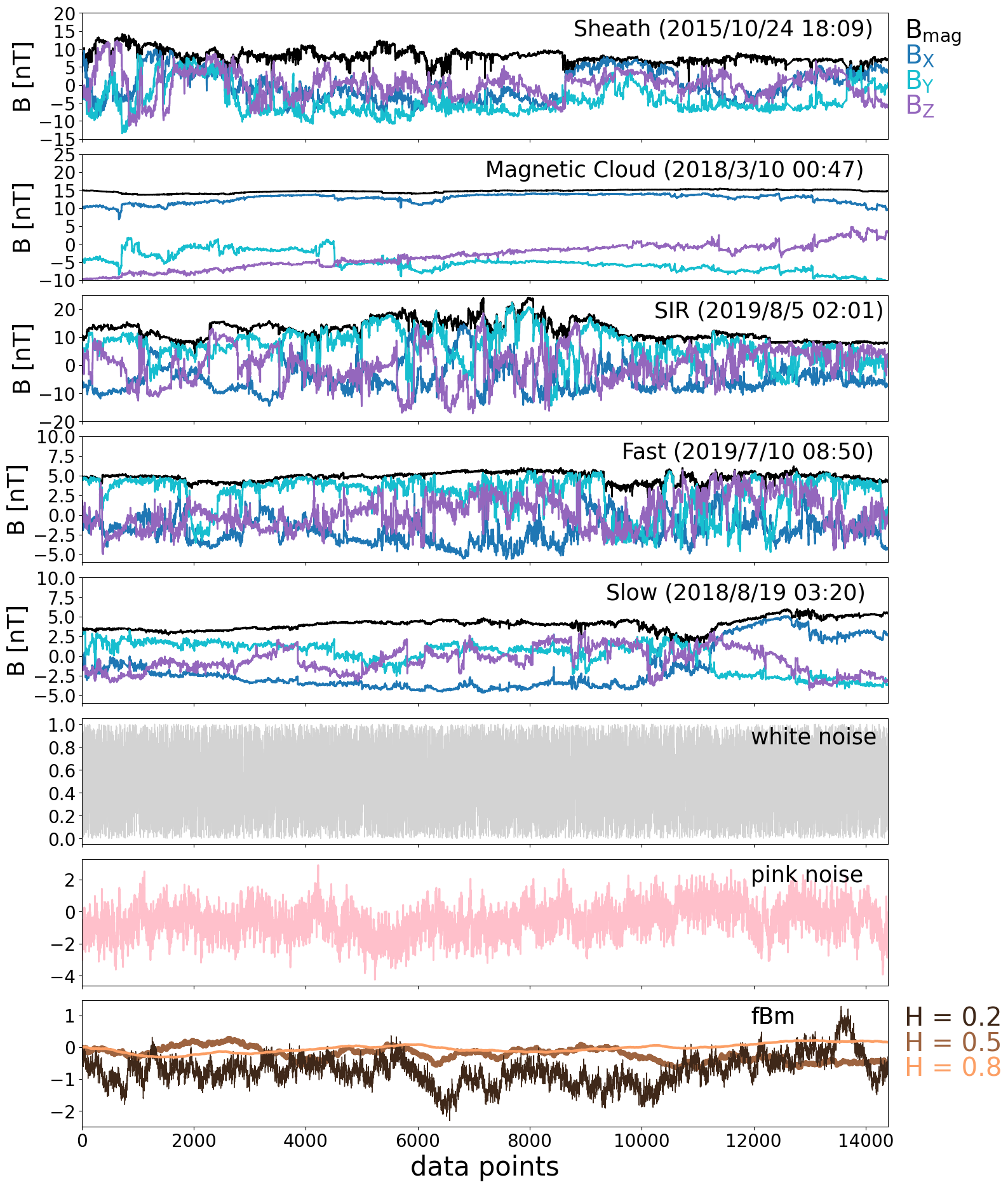}
\caption{Example intervals. The top five panels show the interplanetary magnetic field components in GSE coordinates (blue: $B_X$, cyan: $B_Y$, purple: $B_Z$) and the field magnitude (black). The data shown is 3 s data from the \textit{Wind} spacecraft. The three bottom panels show white noise, pink noise and Brownian motion for the same sample size (14,400 samples) as the solar wind intervals. Brownian motion is shown for three different values of the Hurst exponent, H$=0.2$ (persistent), H$=0.5$ (classical Brownian motion) and H$=0.8$ (trend-reverting).}
\label{fig:example}
\end{figure}

Examples of solar wind data during each solar wind type are presented in Figure \ref{fig:example}. The data series are shown only during the analyzed intervals and boundaries such as the shock and the ICME leading edge for the sheath are thus not included in the plots. It can be seen that the magnetic cloud data series differs considerably from the other four solar wind types displayed. It exhibits smooth variations of all the three field components and a steady magnetic field magnitude. SIRs and sheaths exhibit the most variations in their properties and the field magnitude. The fast wind interval is characterized by the uniform presence of large-amplitude fluctuations that are typically taken to represent anti-sunward propagating Alfv\'{e}n waves. The three bottom panels of Figure~\ref{fig:example} show white noise, pink noise and Brownian motion. These data series were created using the same sample length as the solar wind series above, i.e. 14,400 samples. The white and pink noise were created using a publicly available Python codes. White noise is a maximally random series where different frequencies have equal intensities. Its power spectral density is constant and gives a spectral slope of zero, i.e. for a $f^{-\alpha}$ power law, $\alpha = 0$. Pink noise is associated with the $f^{-1}$ spectrum; compared to white noise, it has relatively more power at low than high frequencies. In the solar wind, the $f^{-1}$ spectral range is often interpreted as the `energy containing' range, found at large scales where energy is believed to be injected in the system before cascading down the turbulent inertial range \citep[e.g.,][]{Bruno2013}. 

In the bottom panel, Brownian motion is shown for three values of the Hurst exponent. The Hurst exponent, $H$, is used to characterize the memory and correlations in the time series \citep[e.g.,][]{Ruzmaikin1994}. The exponent $H = 0.5$ describes the Brownian random walk (also called brown noise or classical Brownian motion) where the mean-squared distance from the starting point of the walk increases with time. Such a time series is uncorrelated in the sense that the steps in the random walk are independent of each other or, in other words, that the auto-correlation of the time series is zero. 

When $H \neq 0.5$ the process is called fractional Brownian motion (fBm). In such cases the increments in the random walk are not independent. The light brown curve in Figure \ref{fig:example} shows the case with $H = 0.8$. When $H > 0.5$ the time series is said to be persistent and exhibit long-term memory or long auto-correlation. The distance from the starting point in the random walk increases faster than in the case of classical Brownian motion. The increasing (decreasing) value is followed by an increase (decrease) and the entropy is lower. The dark brown curve shows the case where  $H = 0.2$. A time series with Hurst exponent $< 0.5$  is  said to be mean-reverting or anti-persistent, i.e. an increase (decrease) is followed by a decrease (increase). Such short-memory series is unpredictable and has higher entropy and lower complexity when compared to brown noise or the case with $H > 0.5$. Now the distance from the starting point increases slower than for classical Brownian motion. A visual inspection of the solar wind time series reveals that the slow solar wind and magnetic cloud time series are the most consistent with the larger Hurst exponents while the other intervals appear more like the short-memory fBm and pink noise. 

\subsection{Ordinal patterns}
\label{sec:ordinal_patterns}

\begin{figure*}[ht]
\centering
\includegraphics[width=0.9\linewidth]{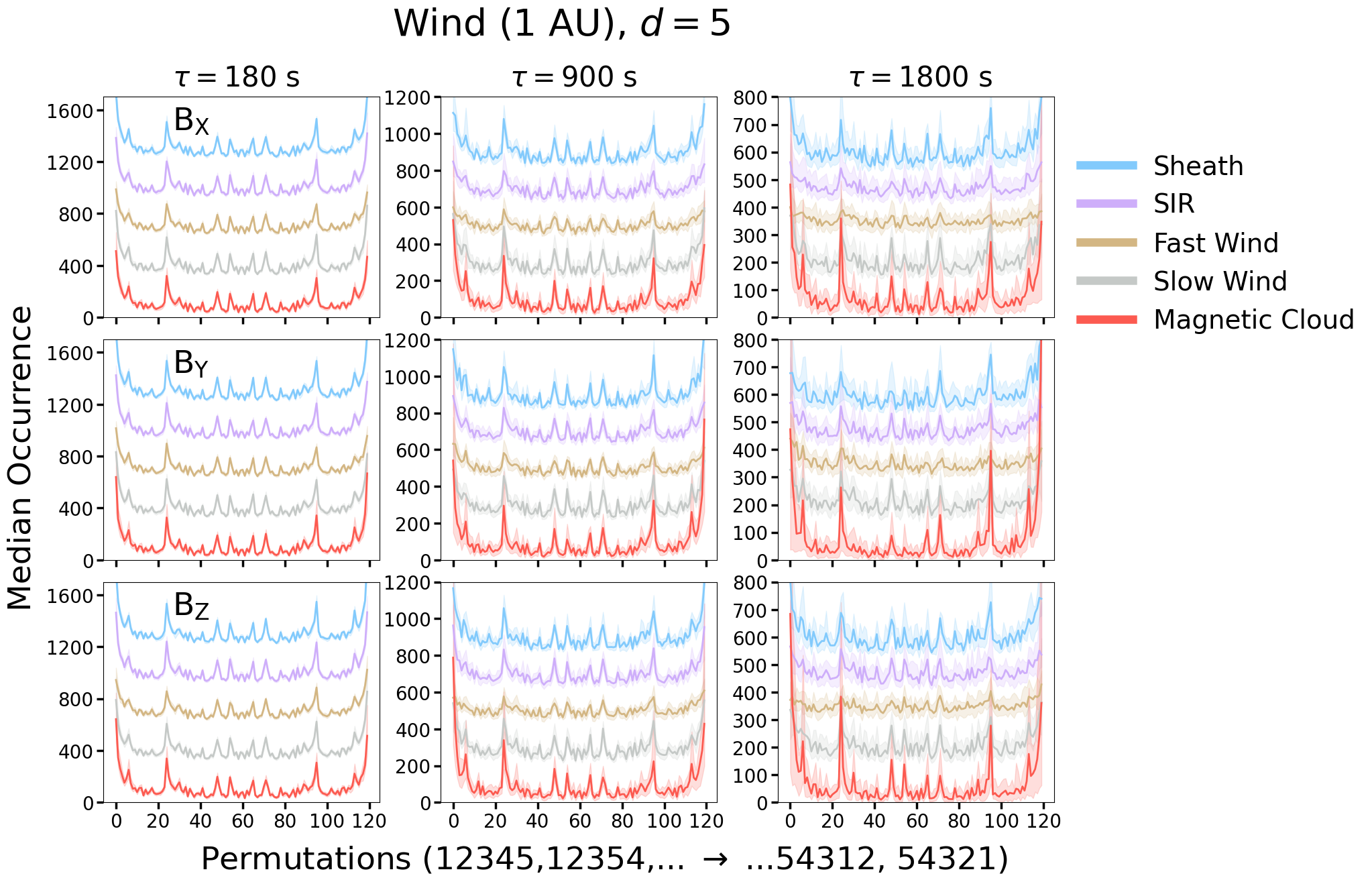}
\caption{Median occurrence of permutations observed in different solar wind categories for $d=5$ for $\tau=$~180, 900 and 1800~s, i.e. for subsampling rates $r=$~60, 300 and 600) for the three GSE magnetic field components. Note that all curves except the magnetic cloud curves have been shifted upwards by a fixed amount by 300 for $\tau=$~180 s, 200 for $\tau=$~900 s and 130 for $\tau=$~1800 s) to aid comparison. Shaded areas show the interquartile range.}
\label{fig:permutations}
\end{figure*}

We first investigate the occurrence of ordinal patterns in time series of the three GSE magnetic field components sampled across all events within the different solar wind categories. Figure \ref{fig:permutations} shows the distributions of median occurrence of permutations for three time lags $\tau=$~180, 600 and 1800 s (corresponding to sub-sampling rates $r=$ 60, 300, and 900) and $d=5$, i.e., each ordinal pattern has five elements. Note that a fixed number has been added to all curves except `Magnetic Cloud' to aid comparison between different categories (see the figure caption for details). The shaded areas show the interquartile ranges. For smallest timescales shown ($\tau=~180$~s), both the lower and upper quartiles are very close to the median while for the largest $\tau=$~1800~s values are more spread.

A clear trend visible in Figure \ref{fig:permutations} is that the permutations with several monotonically increasing and decreasing numbers (12345, 54321, 12354, 21345, etc.) are most abundant, in particular for $\tau=$~180~s for all investigated solar wind categories and magnetic field components.  As the timescale increases such peaks become weaker for the fast wind in particular. For the magnetic clouds and also for sheaths the peaks become even more pronounced. Another striking feature in Figure \ref{fig:permutations} is that for $\tau = $ 180 and 900 s in particular, the peaks and dips in the occurrence of permutations are highly correlated  across all investigated solar wind categories and magnetic field components.  The magnetic cloud category stands out as the one having certain permutations dominating also at the largest scales.

\subsection{Entropy and complexity of fluctuations}
\label{sec:entropy_complexity}

\begin{figure*}[ht]
\centering
\includegraphics[width=0.9\linewidth]{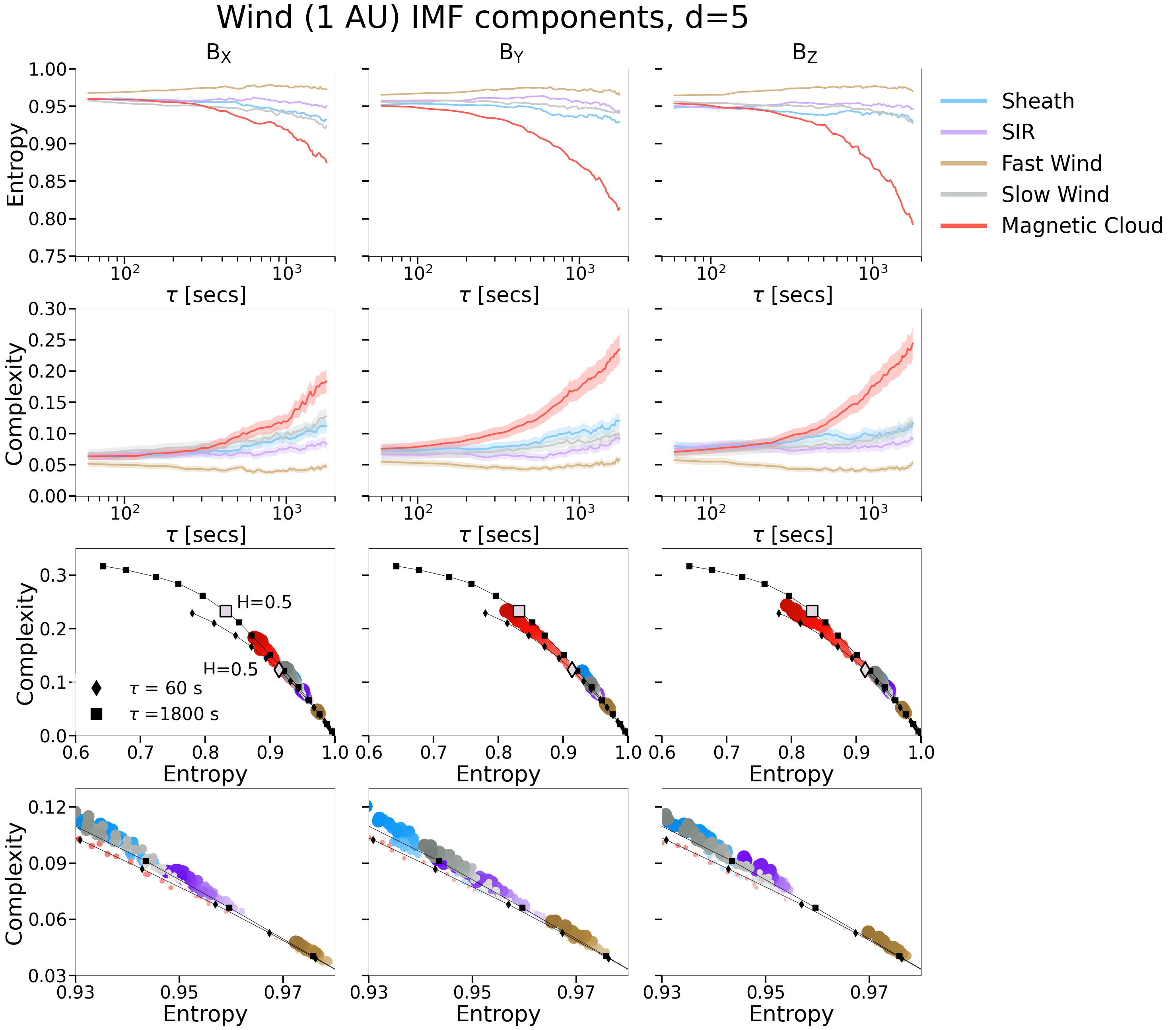}
\caption{Entropy (first-row panels) and complexity (second-row panels) as a function of $\tau$ for $d=5$, for the GSE magnetic field components (panels left to right) in the different solar wind categories (colour-coded lines). Shaded areas in the bottom panels give the uncertainty ranges estimated using the permutation occupation number approach. The third and fourth-row panels gives scatter plots of average entropy and complexity for the different solar wind categories from $\tau =$~60 to 1800 s in steps of 20 s. Time lag increases with the size and darkness of the marker circles. Two curves with black diamonds and squares show fractional Brownian motion for sub-sampling rates $r=20$ and $r=600$, respectively. The grey square and diamond markers show the curves at Hurst exponent 0.5.}
\label{fig:SC_tau}
\end{figure*}

The (normalized) permutation entropy and complexity as a function of time lag $\tau$ (i.e. $r \Delta t$) are investigated in the two top panels of Figure \ref{fig:SC_tau} for $d=5$. The time $\tau$ is increased from 60~s ($r=20$) to 1800~s ($r=600$) in steps of 60~s ($r=20$).
The uncertainty ranges indicated in the complexity panels are estimated from the average permutation occupation number $\sqrt{d!/(N-(d-1) r)}$ \citep[e.g.,][]{Weygand2019,Good2020a}. The uncertainties increase with increasing time lag (i.e. with increasing $r$) as the total number of permutations extracted from the time series decreases slightly with decreasing $\tau$ ($r$).

The top panels of Figure \ref{fig:SC_tau} show 
that the entropies in the fast wind, slow wind, sheath and SIR categories show no or very weak dependencies on $\tau$, with entropy weakly reducing at larger $\tau$ in some components for the latter three categories. The magnetic cloud entropy, in contrast, shows a significant $\tau$ dependence, falling strongly in all three components at large $\tau$.
 For the fast wind  the entropy in turn consistently increases very weakly with $\tau$ before flattening. These trends are similar for all magnetic field components. The key difference is that for magnetic clouds $B_Y$ and $B_Z$ components decrease to considerably lower entropies than $B_X$ at largest $\tau$. The values of entropy also differ between different solar wind categories at $\tau \gtrsim 300$~s. The fast wind consistently has the highest entropy and magnetic clouds the lowest entropy across all three components, with the other categories at intermediate values. 

 The complexity (second-row panels) broadly mirrors the entropy trends: with increasing $\tau$, complexity is approximately invariant in the fast wind, increases weakly in SIRs, sheaths and slow wind, and increases significantly in magnetic clouds. The relatively low entropy and high complexity in the magnetic clouds at large $\tau$ reflects their coherent, ordered structure at large scales, while the high entropy / low complexity of the fast wind reflects its unstructured, stochastic nature at all of the scales we have considered here. Again, the key difference between the components is for magnetic clouds. For the $B_Y$ and $B_Z$ components the complexity increases to larger values than for $B_X$ at largest $\tau$.

The third row of Figure \ref{fig:SC_tau} shows scatter plots of the average entropy and complexity values for the different solar wind categories. The time lags are shown from $\tau =$~60~s (lighter and smaller circles) to 1800~s (darker and larger circles). The bottom row of Figure \ref{fig:SC_tau} magnifies the high-entropy, low-complexity corner of the plots. The curves are shown for fBm with the Hurst exponent running from 0.05 to 0.75 in steps of 0.05, and for both time lags $\tau=60$~s and $\tau=1800$~s (sub-sampling rates $r=20$ and $r=600$). These plots show that the averaged values from the solar wind time series follow closely the fBm curves. In general, the averaged data points move towards the higher Hurst exponent ends of the fBm curves with increasing $\tau$ (increasing sub-sampling rates). The clearest exception is the fast wind. For the fast wind data points  are clustered at the bottom-right corner and its data points exhibit higher entropies and lower complexities at the smaller smaller $\tau$. We also note that for most cases the data points for larger time lags are a bit above the fBm curve indicating a higher complexity. 

\subsection{Complexity--entropy map}
\label{sec:CH_map}

In this section we investigate how the solar wind data series are placed onto the complexity--entropy plane, where the vertical axis shows complexity and the horizontal axis entropy \citep[see e.g.][]{Weck2015, Weygand2019}. Highly stochastic fluctuations are represented by white and pink noise, which appears in the very bottom-right part of the plane with entropy $\sim 1$ and complexity $\sim 0$. Chaotic fluctuations have entropies between $\sim 0.45 - 0.70$ and complexities close to the maximum complexity curve \citep[e.g.,][]{Zanin2021}. Periodic fluctuations (e.g. sinusoidal functions) would fall onto the lower left part of the plane (not shown). They have low entropies $\sim 0 - 0.50$ and, while their complexities follow the maximum complexity curve, they do not attain the peak values. Given that differences between the GSE components were found to be relatively small in section~\ref{sec:ordinal_patterns}, we here choose to investigate $B_Z$ only. $B_Z$ is also the IMF component that has most interest for geoeffectivity \citep[e.g.,][]{Kilpua2017b}.

\begin{figure*}
\captionsetup[subfigure]{labelformat=empty}
\centering
   \begin{subfigure}[b]{0.80\linewidth}
   \includegraphics[width=0.99\linewidth]{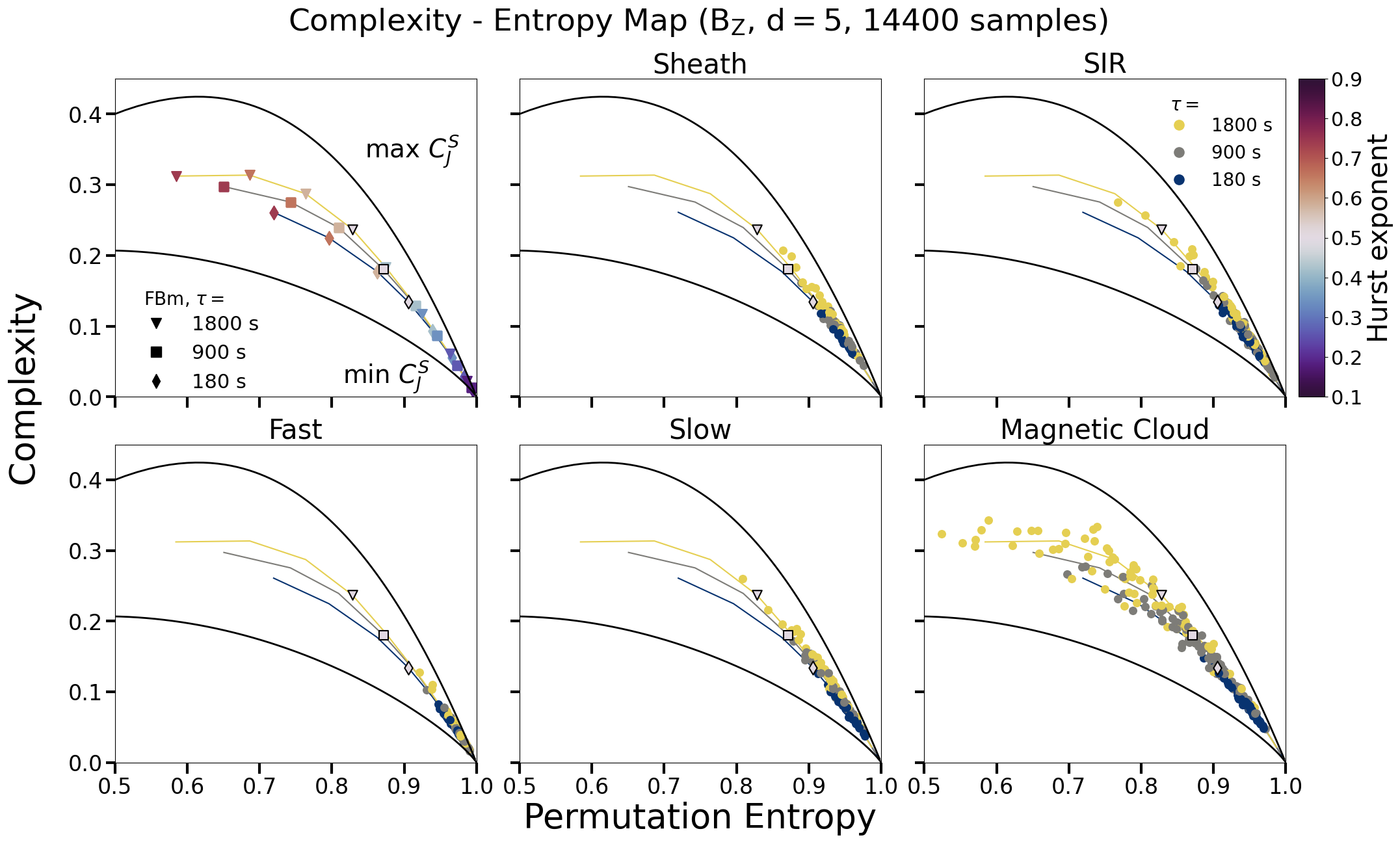}
   \caption{}
\end{subfigure}
   %\\
\begin{subfigure}[b]{0.80\linewidth}
   \includegraphics[width=0.99\linewidth]{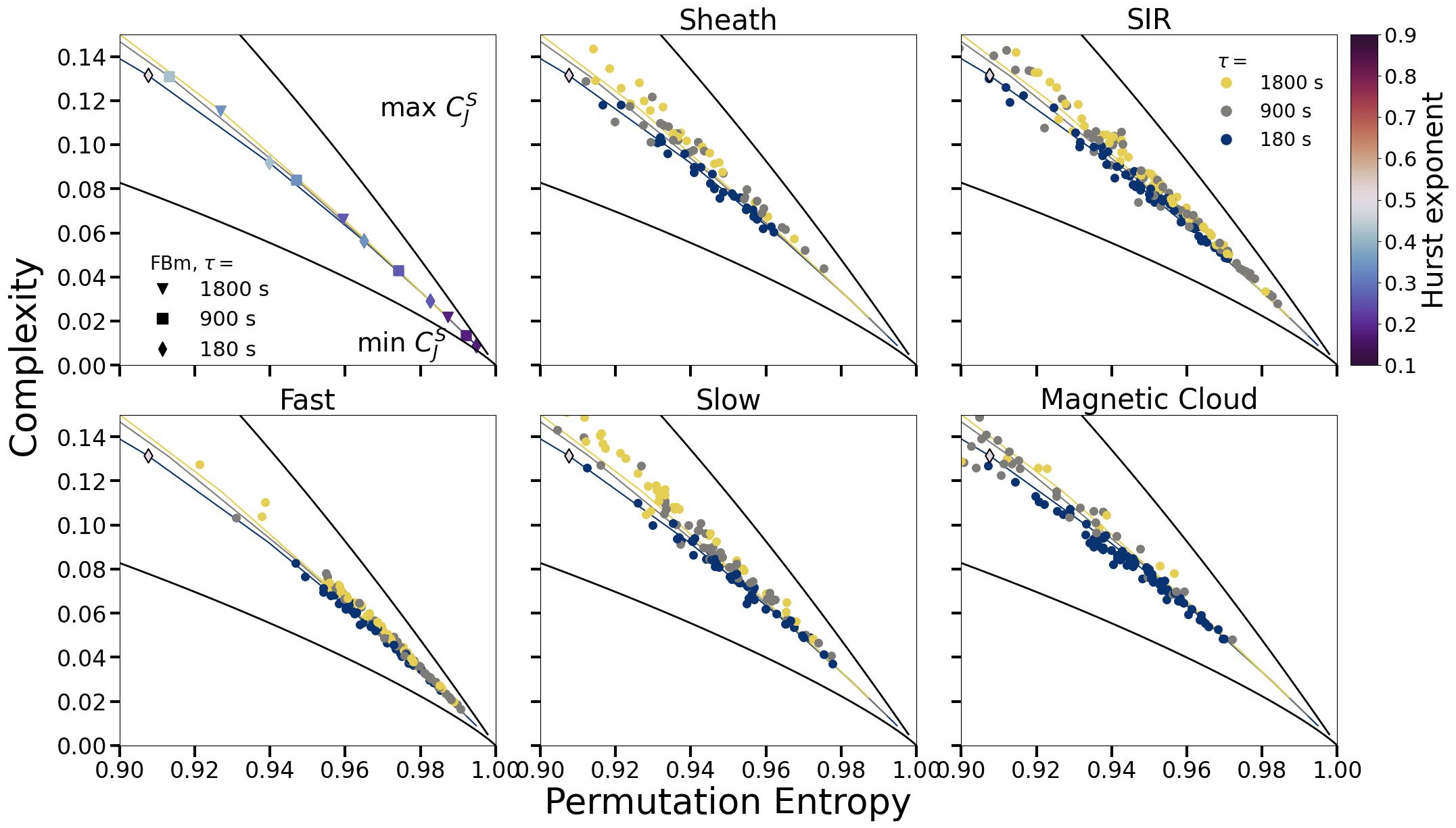}
   \caption{}
\end{subfigure}
\caption{Complexity--entropy maps showing Jensen-Shannon complexity in $B_Z$ plotted against normalized permutation entropy for $d=5$ and sub-sampling rates ranging from $r=20$ to 600 in steps of 60 ($\tau =60$ to 1800~s in steps of 60~s). The dark blue points show values for $r=20$, with point colour darkening to yellow with increasing $r$. In the top left panel the grey square and diamond markers show the fractional Brownian motion calculated with $r=20$ and  $r=600$, respectively, for Hurst exponents from 0.1 to 0.8. The markers repeated in all panels with black edges indicate $H=0.5$. Minimum and maximum complexity curves are shown in black.}
   \label{fig:CS_map2}
\end{figure*}

Figure~\ref{fig:CS_map2} shows the distribution of computed values for different solar wind categories in the complexity--entropy plane for time lags $\tau=$, 180, 900, and 1800~s  (subsampling rates $r =$, 60, 300 900). The dark blue-coloured dots correspond to $\tau=180$~s, gray dots $\tau=900$~s and the yellow dots $\tau=1800$~s. The fBm points for varying Hurst exponents are also shown in the figure, for three investigated time lags / subsampling rates. The two black curves show the minimum and maximum complexity curves. 

Magnetic clouds clearly exhibit the widest spread of data points in the complexity--entropy map. Their entropies reach to the H$=0.8$ markers, and complexities of about 0.35. While data points for $\tau=$180~s fall onto the fBm curve, a significant fraction of data points deviates from the fBm curve for time lags $\tau=$900, and 1800~s. There is a clear trend that the data points move to lower entropies and higher complexities with the increasing time lag.   

The fast wind data points are in turn clustered at the lower right part of the map, with the majority of them having entropies $\gtrsim 0.96$. For the slow wind and compressive structures (i.e. sheaths and SIRs), the lowest entropy values are $\sim 0.8$. For other solar wind structures  the organization with $\tau$ is less clear than for magnetic clouds. While the data points furthest along the fBm curves are solely related to the largest time lag $\tau=1800$~s, in the right lower corner there is a mixture of data points from all included time lags. In particular for the fast wind and SIRs this region is occupied with data points related to $\tau=900$~s and $\tau=1800$~s.  

\subsection{Hurst exponent}
\label{sec:embed_dim}

\begin{figure}
\captionsetup[subfigure]{labelformat=empty}
\centering
   \begin{subfigure}[b]{1.02\linewidth}
   \includegraphics[width=1.02\linewidth]{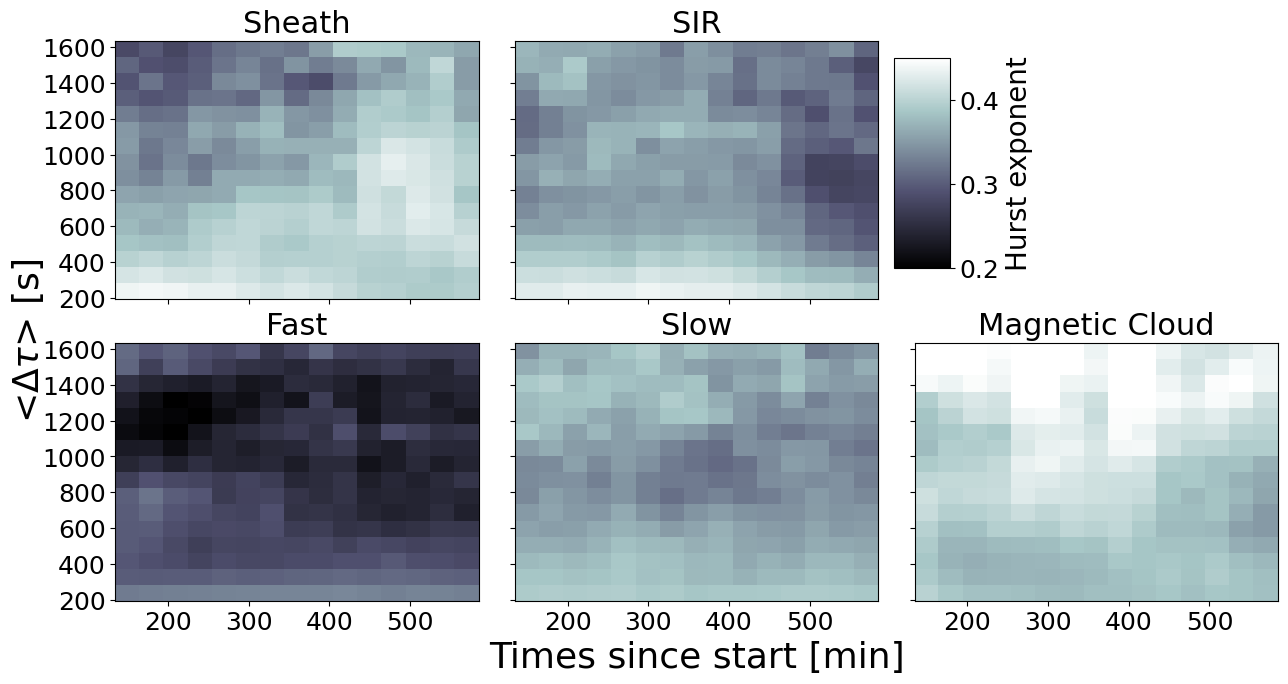}
   \caption{}
\end{subfigure}
   \\
\begin{subfigure}[b]{1.02\linewidth}
   \includegraphics[width=1.02\linewidth]{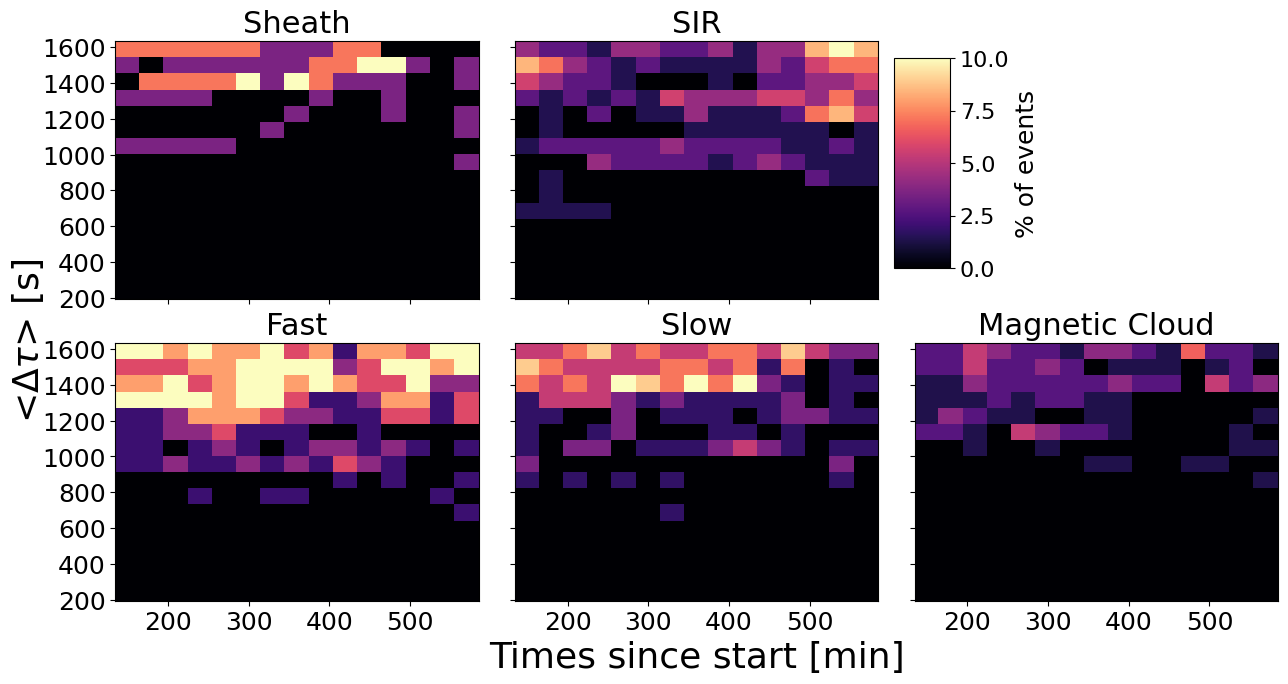}
   \caption{}
\end{subfigure}
\caption{(Top) Colour maps showing the values of the local Hurst exponent as a function of time (given in minutes from the start of the solar wind time interval) and the mid-point of the time-lag interval used to calculate the Hurst exponent (see text for details).  (Bottom) The percentage of the events for which the  the standard error related to the fitting was $> 0.05$. These events were removed from the analysis.}
   \label{fig:hurst1}
\end{figure}

\begin{figure*}[ht]
\centering
\includegraphics[width=0.75\linewidth]{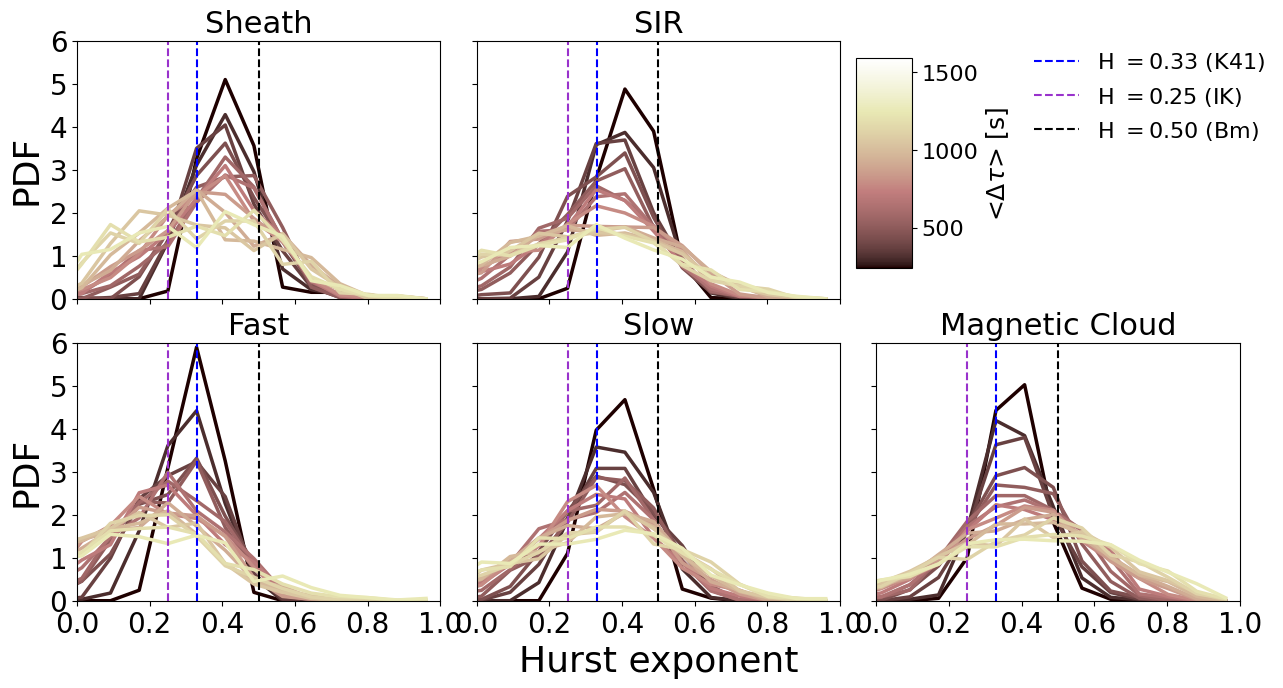}
\caption{Probability density functions (PDFs) of the Hurst exponents calculated over varying time-lag ranges for different solar wind types.}
\label{fig:hurst2}
\end{figure*}

The results of the Hurst exponent analysis are presented in Figures \ref{fig:hurst1} and \ref{fig:hurst2} for the $B_z$ component. The values are averaged across all events.  The top panels of Figure \ref{fig:hurst1} show the colour maps of the  Hurst exponent values as a function of the mid point of the sliding-window $\Delta \tau$-range and the time from the start of the 12~h interval in bins as described above.  The bottom panels of Figure \ref{fig:hurst1} show the percentage of the events when the standard error related to the fitting was $> 0.05$, i.e. the percentage of the removed points. At smallest scales the fitting was good for all events, while at the largest scales up to $\sim 10$ \% had the standard error exceeding our threshold. 

The different solar wind types have some clear differences in their Hurst exponent values and show temporal variations in time. The fast solar wind has clearly the lowest Hurst exponents from the investigated solar wind types, being $\sim 0.20$ or below at the front part of the fast stream at the larger scales.  The low Hurst exponents indicate anti-persistent properties of the time series. In addition, the beginning of the sheaths, i.e. the region following the shock, and the end part of the SIR also have low Hurst exponents. In terms of spectral slopes, values below the Kraichnan scaling (corresponding $H < 0.25$), could  stem from the inclusion of the part of the $f^{-1}$ range, but as discussed in Section 2.3, the connection between the Hurst exponent and spectral slopes is affected by solar wind typically exhibiting multifractality. The inclusion of part of the $f^{-1}$ range could also explain that fast wind had largest percentage of cases with poor fit. The low-frequency break point for the fast wind occurs at relatively high frequencies (at 1~au frequencies corresponding a few hours) while in the slow solar wind, it is found only when long enough (several days) time series are used \citep[e.g.,][]{Bruno2013,Chen2020}. 

The largest average Hurst exponent values are found for the sheaths and in particular for magnetic clouds. For magnetic clouds the largest exponents are clustered at the largest $ \Delta \tau$ and at the mid-part of the cloud.  For sheath, SIR and fast wind the Hurst exponents tend to decrease with increasing timescale while for the magnetic clouds the trend is indeed the opposite.  Figure \ref{fig:hurst1} also reveals a strong locality in Hurst exponents for sheaths and SIRs. Sheaths have clearly the largest Hurst exponents at their end parts, i.e. close to the leading edge of the driving CME ejecta than close to the shock. For SIRs the Hurst exponents in turn get smaller close to the end part of the SIR. The slow wind in turn does not show any obvious trends in the Hurst exponents with time scale or with the investigated interval.

Probability density functions (PDFs) of the Hurst exponent for varying $\Delta \tau$ are shown in Figure \ref{fig:hurst2}. The PDFs here include values combined over the whole 12~h intervals. Firstly, for all investigated cases the PDFs become considerably flatter, i.e., the values are spread over a much broader range with increasing timescale. Secondly, the majority of the Hurst exponent values are in the anti-persistent regime ($H < 0.5$). For the fast wind a strong peak is present at the Hurst exponent values $H=\sim 0.33$ for smaller $\Delta \tau$ that moves to $H=\sim 0.25$ with increasing $\Delta \tau$. In case of monofractal time series,  these peaks represents the Kolmogorov and  Kraichnan scalings, respectively.  

The slow wind PDFs  peak at slightly larger Hurst exponent values and have heavier tail towards large values than for the fast wind. The relative order in the Hurst exponents for the fast and slow wind is in agreement with their spectral slopes 
from several previous studies \citep[e.g.,][]{Teodorescu2015,Borovsky2019,Yordanova2009JGR}, i.e. the fast wind is known to have in general shallower slopes than the slow wind. The sheath and SIR distributions are overall quite similar to the slow wind PDFs. For magnetic clouds the PDFs extent considerably to the persistent $H > 0.5$ regime, and in particularly so for the larger $\Delta \tau$.

\section{Discussion}
\label{sec:discussion}

We first studied the occurrence of ordinal patterns, i.e. permutations in different solar wind types (fast and slow wind, magnetic clouds, SIRs and ICME-driven sheaths). The ordinal patters were constructed here from the set of five data points separated by a time lag $\tau$/sub-sampling rate $r$. We found that their occurrences were very similar in particular for timescale of 180~s and 900~s with certain permutations showing distinct peaks for all solar wind types. The consistent occurrence of certain permutations suggests the presence of coherent structures or waves in the data. 

At the smallest scale 180~s for all investigated solar wind types, we also found a clear dominance of ordinal patterns with a small change between the subsequent values, i.e. implying the absence of sudden and large changes of the magnetic field. Such peaks persisted for sheaths, slow wind and in particular for magnetic clouds also for the largest time lag included in the analysis (1800~s, i.e. 30 minutes), while for the fast wind peaks already got less dominant at the time lag $900~$s. The reason is that at the largest timescales magnetic clouds become increasingly sensitive to the coherent flux rope rotation. The persistence of peaks at large scales for sheath regions and slow wind suggests some coherent global structuring also for these solar wind types. 

The entropy and complexity values between different solar wind types were found to be very similar at the smallest timescales/sub-sampling rates  up to $\tau \sim 300$~s (subsampling rate $r \sim 100$). This suggests a uniformity in the physical processes operating at smaller scales, independent of the large-scale structure of the solar wind. The entropy and complexity remained also relatively stable as a function of $\tau$ for all other solar wind types except for magnetic clouds. For magnetic clouds the entropy values strongly decreased and complexity values strongly increased with increasing $\tau$.

The entropy being approximately constant across  the $\tau$ range with a slight increase towards the largest timescales is a signature of stochastic fluctuations \citep[e.g.,][]{Osmane2019}. This trend was identified here in particular for the fast wind that also had throughout the investigated $\tau$ range the highest entropy  and lowest complexity values. These finding imply that the fast wind is the most stochastic in nature from the investigated solar wind types. This finding is consistent with the fast streams having high Alfv\'{e}nicity due to frequent presence of stochastic Alfv\'{e}nic fluctuations propagating through the wind \citep[e.g.,][]{Belcher1971,Bruno2013}.

The above described entropy and complexity trends found for magnetic clouds are in turn in agreement with our understanding of magnetic clouds as ordered structures featuring low magnetic field variability and smooth rotation of the field direction over an interval of one day \citep[e.g.,][]{Klein1982,Kilpua2017a} and as discussed above,  strong bias toward ordinal patterns where the values were steadily increasing or decreasing. The analysis of Ulysses data by \cite{Raath2022} also suggested that part of the low entropy data periods in their study are associated to ordered magnetic clouds. The reason why we found that in magnetic clouds $B_Y$ and $B_Z$ components show higher complexities at larger timescales than $B_X$ is likely reflecting the fact that the large-scale and coherent magnetic field rotation in them occurs predominantly in $B_Y$ and $B_Z$. As mentioned earlier, their internal configuration is a magnetic flux rope. These structures propagate radially from the Sun, and therefore the minimum rotation of the field is in $B_X$.

Our finding that the fast wind had the higher entropy lower complexity than the slow wind is in contrast to \cite{Weygand2019}, but in agreement with \cite{Weck2015}. The reason could be that \cite{Weygand2019} used the \textit{Ulysses} data at larger heliospheric distances while our study and \cite{Weck2015} use \textit{Wind} data gather at 1~au. Fast wind is considered to be dynamically younger, and therefore expected to present less evolved turbulence (i.e., be less stochastic) than slow wind \citep[e.g.,][]{Weygand2019}. However, it could be that, at least near 1~au, slow wind has considerably more coherent structures, while the fast wind is permeated by Alfv\'{e}nic fluctuations which are inherently stochastic in nature. The intermittency of the fast wind is in fact known to increase with heliospheric distance from the Sun, while for the slow wind it remains approximately the same \citep[e.g.,][]{Marsch1993}. We also note that \cite{Weygand2019} found complexities and entropies in interplanetary CMEs (ICMEs) to be similar to that of the slow wind, while in our study magnetic clouds had distinctly different values at larger time lags. We note that in addition including larger time lags, we separated sheaths from the ejecta and include only magnetic clouds that are the most coherent subset of ICMEs. \cite{Weygand2019} included ICMEs as a whole and they also included all ICMEs, not only magnetic clouds.

The placement of data points in the complexity - entropy plane indicates that for most cases solar wind fluctuations follow relatively closely the fractional Brownian motion (fBm) curves. At the smallest scales the values fall exactly on the fBm curve, while at larger scales there are some deviations, mostly above the fBm curve. This suggest higher ordering that likely stems from the presence of coherent structures such as current sheets, small-scales flux ropes, and magnetic holes. These findings are consistent with previous studies finding solar wind fluctuations being stochastic in nature, i.e the absence of chaotic or periodic fluctuations\cite[e.g.,][]{Weck2015,Weygand2019,Good2020a,Kilpua2022}.  It is however possible that if larger timescales would have been included, the most coherent magnetic clouds would fall into the periodic domain of the complexity - entropy plane as the time series of their magnetic field components resemble that of a half wave.  

The fast wind exhibits least spread in their data points in the complexity - entropy plane and they are clustered closest to the lower right part of the map close to the region where pink or white noise is located. This is in agreement with the previously discussed results of the entropy and complexity analysis suggesting that the fast wind is highly stochastic in nature. 

The widest spread of values in the complexity - entropy plane was observed for magnetic clouds for the largest time lags. This could partly stem for a large variety of magnetic cloud structures observed in interplanetary space, from those exhibiting distinctly smooth rotation to cases where considerable distortion is present \citep[e.g.,][]{Kilpua2017a}. However, for magnetic clouds also, the data points at 180~s were clustered close onto the fBm curve, consistent with our previous suggestion of uniformity of the processes at smaller scales in all solar wind types.

The relative similarity in fluctuation properties and placement in the complexity - entropy plane for the slow wind and the large-scale compressive structures (sheaths and SIRs) could stem from the latter mostly consisting of the processed slow wind. Although according to previous studies fluctuation properties change from the upstream to downstream at ICME-driven shocks and some new fluctuations are created, they do not appear to reset the turbulence in a similar manner to planetary bow shocks \citep{Kilpua2021Fr}.  

Both the entropy-complexity maps analysis and PDFs of Hurst exponents derived from the first-order structure function analysis showed a wide spread of Hurst exponents for the investigated time series at the larger time lags. It is also interesting to note that for all other investigated solar wind types except for magnetic clouds most data points reaching the lowest right corner of the complexity - entropy map,  i.e. data points associated with the lowest Hurst exponents, were related to $\tau =$ 900~s and 1800~s. In turn, all data points that were at the highest Hurst exponent values along the fBm curve were related almost solely to the largest time lag 1800~s. 

As discussed in Section 3.3 the Hurst exponent expresses whether fluctuations are persistent, random or anti-persistent. Both the complexity - entropy maps and the  Hurst exponents obtained from the structure function analysis indicate dominantly anti-persistent fluctuations (H $< 0.5$) for all investigated solar wind types and timescales. 

The exponents extending to the persistent regime (H $>0.5$) were identified mostly in magnetic clouds and for the largest time-scales.  Visually (Figure \ref{fig:example}) time series of the magnetic field components extracted during magnetic clouds resemble persistent time series with large Hurst exponent. We note that a significant fraction of magnetic clouds had however Hurst exponents close to $H=0.5$, i.e. being indicative of uncorrelated random walk. Using the relation $\alpha = 2H + 1$ this correspond to the inertial range spectral slope $-2$, i.e. considerably steeper than Kolmogorov's. Recently, \citet{Good2023} showed that magnetic clouds in the inner heliosphere can indeed exhibit slopes as steep as -2 at large scales. Rather than a property of the fluctuations, \citet{Good2023} interpret these steep slopes as a feature of the background magnetic structure, with rotation of the global flux rope field in the clouds adding power to the spectra at large scales. We however again caution drawing strong conclusions on the relation between the Hurst exponent and spectral slopes due to effect of multifractality.

We found significant locality for the Hurst exponents, in particular for compressive SIRs and sheath structures. For sheaths the trailing part had larger Hurst exponents than the front part, while the for SIRs the trend is vice versa. We note that \citep{Kilpua2021Fr} found also differences in spectral slopes between the leading and trailing parts of the sheath. The larger Hurst exponents in the back of the sheath could stem from highly fluctuating and more compressive fields near the flux rope's leading edge. The draping of the field lines around the flux rope, reconnection and depletion regions can lead to current sheaths and discontinuities. For magnetic clouds the larger Hurst exponents at mid part of the cloud could stem from this region representing the least disturbed part of the structure. The boundaries of magnetic clouds are often distorted by their interaction with the ambient solar wind \citep[e.g.,][]{Kilpua2013}. We note that \cite{Balasis2006} found for a change in the Hurst exponents for geomagnetic time series from anti-persistent to persistent properties preceding intense magnetospheric storms, i.e. before the occurrence of an extreme event. In our work, we analysed solar wind time series only within the large-scale structures, but it will be an interesting future work to extent the study to analyse the temporal variations in the scaling properties of continuous solar wind time series over longer periods.

\conclusions

In this work, we have characterized time series sampled in fast and slow wind, magnetic clouds, CME-driven sheaths and SIRs. The time series were analyzed by estimating their permutation entropy, Jensen-Shannon complexity and Hurst exponent from the first-order structure function. The results reflect different dynamical processes behind the generation and evolution of the solar wind structures and different behaviours with varying time scale. At small scales, all of the solar wind types show similar occurrence frequency of ordinal patterns, entropy and complexity values, while clear differences are evident at large scales. All solar wind types except magnetic clouds at largest scales follow relatively closely fractional Brownian motion (fBm) in the complexity-entropy plane but are partly located at different parts of the time-scale dependent fBm curve. The fast wind and magnetic clouds stood out as having the most distinct fluctuation characteristics, while the slow wind and compressive structures (SIRs and sheaths) resembled more closely each other. We also found a significant non-locality in Hurst exponents, in particular for sheaths and SIRs.

 This study demonstrates that permutation entropy and complexity analysis is a useful tool for investigating the solar wind and its large-scale structures. The analysis can help to explore their internal processes, and how these internal processes relate to the local fluctuation properties. In addition, the complexity-entropy analysis could  reveal the occurrence of mesoscale structures in a statistical sense in space plasmas at different scales as they are expected to add more structures to the data, and hence leading to higher complexity. Observations from the fleet of recent launched spacecraft (Solar Orbiter, Parker Solar Probe and BepiColombo) are also expected to yield important information on variations with heliospheric distance.

\dataavailability{The solar wind data used in this study are available from the NASA Goddard Space Flight Center Coordinated Data Analysis Web (CDAWeb; \url{http://cdaweb.gsfc.nasa.gov/}). The Wind ICME list is available at \url{https://wind.nasa.gov/ICME_catalog/}, the Richardson and ICME list at \url{https://izw1.caltech.edu/ACE/ASC/DATA/level3/icmetable2.htm}) \citep{richardson2010}, and the ACE/WIND SIRs catalog at \url{http://www.srl.caltech.edu/ACE/ASC/DATA/level3/SIR_List_1995_2009_Jian.pdf}. The pink and white noise were generated using the code publicly available at \url{https://github.com/felixpatzelt/colorednoise} and  the Fractional Brownian Motion with the package publicly available at \url{https://pypi.org/project/fbm/}.} 

\authorcontribution{EK performed the data analysis and prepared the figures. All authors have contributed to the writing of the manuscript and interpretation of the results.}

\competinginterests{No competing interests are present.} 

\begin{acknowledgements}
We acknowledge the Finnish Centre of Excellence in Research of Sustainable Space (Academy of Finland grant number 312390). E.K. acknowledges the ERC under the European Union's Horizon 2020 Research and Innovation Programme Project SolMAG 724391. S.G. is supported by Academy of Finland grants 338486 and 346612 (INERTUM). M.A.-L. acknowledges the Emil Aaltonen Foundation for financial support.
\end{acknowledgements}

\bibliographystyle{copernicus}
\bibliography{reference}

\end{document}